\documentclass[leqno]{article}

\usepackage{amssymb,amsmath}
\usepackage{latexsym}

\newcommand{\id}{\mathrm{id}}

\newcommand{\tr}{\mathrm{tr}}
\newcommand{\sign}{\mathrm{sign}}
\newcommand{\cyb}{\mathrm{CYB}}
\newcommand{\mat}{\mathrm{Mat}}
\newcommand{\GL}{\mathrm{GL}}

\newcommand{\im}{\mathrm{Im}}

\newcommand{\sol}{\mathrm{Solv}}

\newcommand{\alt}{\mathcal{A}}
\newcommand{\sym}{\mathcal{S}}
\newcommand{\rt}{\mathcal{R}}
\newcommand{\lie}{\mathcal{L}}

\newcommand{\cg}{\mathfrak{c}}
\newcommand{\ag}{\mathfrak{a}}
\newcommand{\lev}{\mathfrak{l}}
\newcommand{\h}{\mathfrak{h}}
\newcommand{\s}{\mathfrak{s}}
\newcommand{\g}{\mathfrak{g}}
\newcommand{\gl}{\mathfrak{gl}}
\newcommand{\slg}{\mathfrak{sl}}
\newcommand{\su}{\mathfrak{su}}

\newcommand{\F}{\mathbb{F}}
\newcommand{\R}{\mathbb{R}}
\newcommand{\C}{\mathbb{C}}

\begin{document}


\title{Existence of Triangular Lie Bialgebra Structures II}

\author{J\"org Feldvoss\thanks{E-mail address: \tt jfeldvoss@jaguar1.usouthal.edu}\\
{\small Department of Mathematics and Statistics}\\{\small University of South Alabama}\\
{\small Mobile, AL 36688--0002, USA}}

\date{Dedicated to the memory of my father}

\maketitle


\begin{abstract}

\noindent We characterize finite-dimensional Lie algebras over an
arbitrary field of characteristic zero which admit a non-trivial
(quasi-) triangular Lie bialgebra structure.
\medskip

\end{abstract}


\section{Introduction}


In general, a complete classification of all triangular Lie bialgebra structures 
is very difficult. Nevertheless, Belavin and Drinfel'd succeeded in \cite{BD} 
to obtain such a classification for {\it every\/} finite-dimensional {\it simple\/} 
Lie algebra over the complex numbers. The aim of this paper is much more modest 
in asking when there exist {\it non-trivial\/} (quasi-) triangular Lie bialgebra 
structures. Michaelis showed in \cite{Mi2} that the existence of a two-dimensional 
non-abelian subalgebra implies the existence of a non-trivial triangular Lie bialgebra 
structure over any ground field of arbitrary charactristic. In \cite{Fe1} we used a 
slight generalization of the main result of \cite{Mi2} (cf.~also \cite[Section 7]{BD}) 
in order to establish a non-trivial triangular Lie bialgebra structure on almost any 
finite-dimensional Lie algebra over an algebraically closed field of arbitrary 
characteristic. Moreover, we obtained a characterization of those finite-dimensional 
Lie algebras which admit non-trivial (quasi-) triangular Lie bialgebra structures. In 
this paper we extend the results of \cite{Fe1} and \cite{Fe2} to arbitrary ground fields 
of characteristic zero. A crucial result in the first part of this paper is \cite[Theorem 
1]{Fe1} which is not available for non-algebraically closed fields and has to be replaced by 
a more detailed analysis. As in the previous papers, it turns out that with the exception 
of a few cases occurring in dimension three, every finite-dimensional non-abelian Lie 
algebra over an arbitrary field of characteristic zero admits a {\it non-trivial 
triangular\/} Lie bialgebra structure. As a consequence we also obtain that every 
finite-dimensional non-abelian Lie algebra over an arbitrary field of characteristic 
zero has a {\it non-trivial coboundary\/} Lie bialgebra structure. The latter extends 
the main result of \cite{dSm} from the real and complex numbers to arbitrary ground 
fields of characteristic zero.

Let us now describe the contents of the paper in more detail. In Section 2 we introduce 
the necessary notation and prove some preliminary results reducing the existence of 
non-trivial triangular Lie bialgebra structures to three-dimensional Lie algebras or in 
one case showing at least that the derived subalgebra is abelian of dimension at most two. 
The next section is entirely devoted to the existence of non-trivial (quasi-) triangular 
Lie bialgebra structures on three-dimensional {\it simple\/} Lie algebras. It is well-known 
that every three-dimensional simple Lie algebra is the factor algebra of the Lie algebra 
associated to a quaternion algebra modulo its one-dimensional center. Since the Lie bracket
of these so-called quaternionic Lie algebras are explicitly given, one can compute the
solutions of the classical Yang-Baxter equation with invariant symmetric part which generalizes
\cite[Example 1]{Fe1}. This enables us to prove that a three-dimensional simple Lie algebra 
over an arbitrary field $\F$ of characteristic not two admits a non-trivial (quasi-) triangular 
Lie bialgebra structure if and only if it is isomorphic to the split three-dimensional simple 
Lie algebra $\slg_2(\F)$. Moreover, we observe that the classical Yang-Baxter operator for every 
three-dimensional simple Lie algebra over an arbitrary field $\F$ of characteristic not two is
closely related to the determinant. In the last section we finally prove the characterization 
of those finite-dimensional Lie algebras which admit {\it non-trivial (quasi-) triangular\/} 
Lie bialgebra structures and show that every finite-dimensional non-abelian Lie algebra over 
an arbitrary field of characteristic zero admits a {\it non-trivial coboundary\/} Lie bialgebra 
structure.


\section{Preliminaries}


Let $\F$ be a commutative field of arbitrary characteristic. A
{\it Lie coalgebra\/} over $\F$ is a vector space $\cg$ over $\F$
together with a linear transformation $$\delta :
\cg\longrightarrow\cg\otimes\cg,$$ such that
\begin{equation}
\im(\delta)\subseteq\im(\id_\cg-\tau),
\end{equation}
and
\begin{equation}
(\id_\cg+\xi+\xi^2)\circ(\id_\cg\otimes\delta)\circ\delta=0,
\end{equation}
where $\id_\cg$ denotes the identity mapping on $\cg$, $\tau:\cg
\otimes\cg\to\cg\otimes\cg$ denotes the {\it switch mapping\/}
sending $x\otimes y$ to $y\otimes x$ for every $x,y\in\cg$, and
$\xi:\cg\otimes\cg\otimes\cg\to\cg\otimes\cg\otimes\cg$ denotes
the {\it cycle mapping\/} sending $x\otimes y\otimes z$ to
$y\otimes z \otimes x$ for every $x,y,z\in\cg$. The mapping
$\delta$ is called the {\it cobracket\/} of $\cg$, (1) is called
{\it co-anticommutativity\/}, and (2) is called the {\it co-Jacobi
identity\/}. Note that any cobracket on a one-dimensional Lie
coalgebra is the zero mapping since $\im(\id-\tau)=0$. This is dual
to the statement that every bracket on a one-dimensional Lie
algebra is zero. For further information on Lie coalgebras we
refer the reader to \cite{Mi1} and the references given there.

A {\it Lie bialgebra\/} over $\F$ is a vector space $\ag$ over
$\F$ together with linear transformations $[\cdot,\cdot]:\ag
\otimes\ag\to\ag$ and $\delta:\ag\to\ag\otimes\ag$ such that 
$(\ag,[\cdot,\cdot])$ is a Lie algebra, $(\ag,\delta)$ is a Lie
coalgebra, and $\delta$ is a {\it derivation\/} from the Lie
algebra $\ag$ into the $\ag$-module $\ag\otimes\ag$, i.e.,
$$\delta([x,y])=x\cdot\delta(y)-y\cdot\delta(x)\qquad\qquad
\forall~x,y\in\ag,$$ where the tensor product $\ag\otimes\ag$ is
an $\ag$-module via the {\it adjoint diagonal action\/} defined by
$$x\cdot \left(\sum_{j=1}^n a_j\otimes b_j\right):=\sum_{j=1}^n
([x,a_j]\otimes b_j+a_j\otimes [x,b_j])\qquad\forall~x,a_j,b_j
\in\ag$$ (cf.~\cite[Section 1.3A]{CP}). A Lie bialgebra structure
$(\ag,\delta)$ on a Lie algebra $\ag$ is called {\it trivial\/} if
$\delta=0$.

A {\it coboundary Lie bialgebra\/} over $\F$ is a Lie bialgebra
$\ag$ such that the cobracket $\delta$ is an {\it inner
derivation}, i.e., there exists an element $r\in\ag\otimes\ag$
such that $$\delta(x)=x\cdot r\qquad\qquad\qquad\forall~x\in\ag$$
(cf.~\cite[Section 2.1A]{CP}).

In order to describe those tensors $r$ which give rise to a
coboundary Lie bialgebra structure, we will need some more
notation. For $r=\sum_{j=1}^n r_j\otimes r_j^\prime\in\ag
\otimes\ag$ set $$r^{12}:=\sum_{j=1}^n r_j\otimes
r_j^\prime\otimes 1,\quad r^{13}:=\sum_{j=1}^n r_j\otimes 1\otimes
r_j^\prime,\quad r^{23}:=\sum_{j=1}^n 1\otimes r_j\otimes
r_j^\prime,$$ where $1$ denotes the identity element of the
universal enveloping algebra $U(\ag)$ of $\ag$. Note that the
elements $r^{12},r^{13},r^{23}$ are considered as elements of the
associative algebra $U(\ag) \otimes U(\ag)\otimes U(\ag)$ via the
canonical embedding $\ag \hookrightarrow U(\ag)$. Therefore one
can form the commutators given by
\begin{eqnarray*} && [r^{12},r^{13}]=\sum_{i,j=1}^n [r_i,r_j]
\otimes r_i^\prime \otimes r_j^\prime,\\ && [r^{12},r^{23}]=
\sum_{i,j=1}^n r_i\otimes [r_i^\prime,r_j] \otimes r_j^\prime,\\
&& [r^{13},r^{23}]=\sum_{i,j=1}^n r_i\otimes r_j\otimes
[r_i^\prime,r_j^\prime].
\end{eqnarray*}
Then the mapping $$\cyb:\ag\otimes\ag\longrightarrow\ag\otimes
\ag\otimes\ag$$ defined via $$r\longmapsto[r^{12},r^{13}]+
[r^{12},r^{23}]+[r^{13},r^{23}]$$ is called the {\it classical
Yang-Baxter operator\/} for $\ag$. The equation $\cyb(r)=0$ is the
{\it classical Yang-Baxter equation\/} (CYBE) for $\ag$, and a
solution of the CYBE is called a {\it classical $r$-matrix\/} for
$\ag$ (cf.~\cite[Section 2.1B]{CP}).

Assume for the moment that the characteristic of the ground field
$\F$ is not two. For any vector space $V$ over $\F$ and every
natural number $n$ the {\it symmetric group\/} $S_n$ of degree $n$
acts on the $n$-fold tensor power $V^{\otimes n}$ of $V$ via
$$\sigma\cdot(v_1\otimes\dots\otimes v_n):=v_{\sigma(1)}\otimes
\dots\otimes v_{\sigma(n)}\quad\forall~\sigma\in S_n;v_1,\dots,
v_n\in V.$$ The $\F$-linear transformation $\sym_n:V^{\otimes n}\to
V^{\otimes n}$ defined by $t\mapsto\sum_{\sigma\in S_n}\sigma
\cdot t$ is called the {\it symmetrization transformation\/}. The
elements of the image $\im(\sym_n)$ of $\sym_n$ are just the {\it
symmetric\/} $n$-tensors, i.e., elements $t\in V^{\otimes n}$ such
that $\sigma\cdot t=t$ for every $\sigma\in S_n$. Moreover,
$\im(\sym_n)$ is canonically isomorphic to the $n$-th {\it
symmetric power\/} $S^n V$ of $V$. The $\F$-linear transformation
$\alt_n:V^{\otimes n}\to V^{\otimes n}$ defined by $t\mapsto
\sum_{\sigma\in S_n}\sign(\sigma)(\sigma\cdot t)$ is called the
{\it skew-symmetrization\/} (or {\it alternation\/}) {\it
transformation\/}. The elements of the image $\im(\alt_n)$ of $\alt_n$
are just the {\it skew-symmetric\/} $n$-tensors, i.e., elements
$t\in V^{\otimes n}$ such that $\sigma\cdot t=\sign(\sigma)t$ for
every $\sigma\in S_n$. Since $\im(\alt_n)$ is canonically
isomorphic to the $n$-th {\it exterior power\/} $\Lambda^n V$ of
$V$, in the following we will always identify skew-symmetric 
$n$-tensors with elements of $\Lambda^n V$; e.g., we write 
$$v_1\wedge v_2=(\id_V-\tau)(v_1\otimes v_2)=v_1\otimes v_2-v_2\otimes v_1$$ 
in the case $n=2$, where $\tau : V^{\otimes 2}\to V^{\otimes 2}$ is 
given by $v_1\otimes v_2\mapsto v_2\otimes v_1$, and $$v_1\wedge v_2 
\wedge v_3=\sum_{\sigma\in S_3}\sign(\sigma)v_{\sigma(1)}\otimes 
v_{\sigma(2)}\otimes v_{\sigma(3)}$$ in the case $n=3$. As a direct 
consequence of our identifications, we also have that $$V^{\otimes 2}
=S^2V\oplus\Lambda^2 V.$$

If $\ag$ is a Lie algebra and $M$ is an $\ag$-module, then the set
of $\ag$-{\it invariant elements\/} of $M$ is defined by
$$M^\ag:=\{m\in M\mid a\cdot m=0\quad\forall~a\in\ag\}.$$
Let $r\in\ag\otimes\ag$ and define $\delta_r(x):=x\cdot r$ for
every $x\in\ag$. Obviously, a coboundary Lie bialgebra structure
$(\ag,\delta_r)$ on $\ag$ is {\it trivial\/} if and only if
$r\in(\ag\otimes\ag)^\ag$. Moreover, Drinfel'd observed that
$\delta_r$ defines a Lie bialgebra structure on $\ag$ if and only
if $r+\tau(r)\in (\ag\otimes\ag)^\ag$ and
$\cyb(r)\in(\ag\otimes\ag\otimes \ag)^\ag$ (see \cite[Section 4,
p.~804]{D} or \cite[Proposition 2.1.2]{CP}). In particular, every
solution $r$ of the CYBE satisfying $r+\tau(r)\in
(\ag\otimes\ag)^\ag$ gives rise to a coboundary Lie bialgebra
structure on $\ag$. Following Drinfel'd such Lie bialgebra
structures are called {\it quasi-triangular}, and quasi-triangular
Lie bialgebra structures arising from skew-symmetric classical
$r$-matrices are called {\it triangular\/}.

In \cite[Example 2]{Fe1} we already observed that the three-dimensional 
{\it Heisenberg algebra\/} does {\it not\/} admit {\it any\/} non-trivial 
triangular Lie bialgebra structure. Let us recall that the (non-abelian) 
{\it nilpotent\/} three-dimensional {\it Heisenberg algebra\/} $$\h_1(\F)
=\F p\oplus\F q\oplus\F\hbar$$ is determined by the so-called {\it
Heisenberg commutation relation\/} $$[p,q]=\hbar.$$

Let us conclude this section by several preliminary results which will be 
used in the proof of the main theorem of this paper. The first lemma is 
already contained in \cite{Fe2} and is an immediate consequence of the 
arguments in the proofs of \cite[Theorem 2 and Theorem 3]{Fe1}.
\bigskip

\noindent {\bf Lemma 1.} {\it Let $\ag$ be a finite-dimensional 
non-abelian Lie algebra over an arbitrary field $\F$ with non-zero 
center. If $\ag$ is not isomorphic to the three-dimensional Heisenberg 
algebra, then $\ag$ admits a non-trivial triangular Lie bialgebra 
structure.}\quad $\Box$
\bigskip

The second lemma follows from a generalization of the proofs of 
\cite[Lemma 4.2 and a part of Lemma 4.1]{dSm} from $\R$ and $\C$ 
to arbitrary ground fields of characteristic zero (for a another 
generalization of the latter see also \cite[Theorem 1]{Fe1}). In 
fact, the proofs in \cite{dSm} remain valid in the more general 
setting.
\bigskip

\noindent {\bf Lemma 2.} {\it Let $\ag$ be a finite-dimensional 
centerless solvable Lie algebra over a field $\F$ of characteristic 
zero. If $[\ag,\ag]$ is non-abelian or $\dim_\F [\ag,\ag]\ge 3$, 
then $\ag$ admits a non-trivial triangular Lie bialgebra structure.}
\quad $\Box$
\bigskip

\noindent {\it Remark.\/} The proof of \cite[Lemma 4.2]{dSm} is even
valid for any field of characteristic $\ne 2$, but the proof of 
\cite[Lemma 4.1]{dSm} uses in an essential way that the ground field
has characteristic zero.
\bigskip

Finally, in the non-solvable case we will need the following result.
\bigskip

\noindent {\bf Lemma 3.} {\it Let $\ag$ be a finite-dimensional 
non-solvable Lie algebra over a field $\F$ of characteristic zero. 
If $\ag$ is not three-dimensional simple, then $\ag$ admits a 
non-trivial triangular Lie bialgebra structure.}
\bigskip

\noindent {\it Proof\/.} Suppose that $\ag$ does not admit any 
non-trivial triangular Lie bialgebra structure. Since the ground 
field is assumed to have characteristic zero, the Levi decomposition 
theorem (see \cite[p.~91]{Jac}) yields the existence of a semisimple 
subalgebra $\lev$ of $\ag$ (a so-called {\it Levi factor\/} of $\ag$) 
such that $\ag$ is the semidirect product of $\lev$ and its solvable 
radical $\sol(\ag)$. Because $\ag$ is not solvable, the levi factor 
$\lev$ is non-zero.

Since $\F$ is infinite, $\lev$ contains a Cartan subalgebra $\h$
(see \cite[Corollary 1.2]{Bar}). Let $\overline{\F}$ denote the algebraic 
closure of the ground field and set $\overline{\lev}:=\lev\otimes_\F 
\overline{\F}$ as well as $\overline{\h}:=\h\otimes_\F \overline{\F}$. 
Then $\overline{\h}$ is also a Cartan subalgebra of $\overline{\lev}$.
According to the structure theory of finite-dimensional semisimple
Lie algebras over algebraically closed fields of characteristic
zero (cf.~\cite[Section IV.1]{Jac}), $\overline{\h}$ is abelian and
$\overline{\lev}$ is the direct sum of $\overline{\h}$ and the
one-dimensional root spaces $\overline{\lev}_\alpha$ with $0\ne
\alpha\in\rt$, where $\rt$ denotes the set of roots of
$\overline{\lev}$ relative to $\overline{\h}$.

Suppose that $\dim_{\overline{\F}}\overline{\h}\ge 2$. Then choose
two linearly independent elements $h,h^\prime\in\h$ and define
$r:=h\wedge h^\prime$. Clearly $\cyb(r)=0$, and it is obvious that 
$\delta_r\ne 0$ if and only if $\overline{\delta_r}:=\delta_r
\otimes\id_{\overline{\F}}\ne 0$. Furthermore, set $\overline{h}:=
h\otimes 1_{\overline{\F}}$ resp.~$\overline{h^\prime}:=h^\prime
\otimes 1_{\overline{\F}}$ and choose $\alpha\in\rt$ with 
$\alpha(\overline{h})\ne 0$. Then for every root vector $X\in
\overline{\lev}_\alpha$ we have $$\overline{\delta_r}(X)=
\alpha(\overline{h})\overline{h^\prime}\wedge X-\alpha(
\overline{h^\prime})\overline{h}\wedge X\ne 0.$$

Hence we can assume that $\dim_{\overline{\F}}\overline{\h}=1$,
i.e., $\overline{\lev}\cong\slg_2(\overline{\F})$. In particular,
$\lev$ is three-dimensional simple.

Suppose now that $\s:=\sol(\ag)\ne 0$. Without loss of generality we can 
assume that $\overline{\h}=\overline{\F}\,\overline{h}$ with $\overline{h}:=
h\otimes 1_{\overline{\F}}$. Then it is well-known from the representation
theory of $\slg_2(\overline{\F})$ that either the weight space $(\s\otimes_\F
\overline{\F})_0\cong\s_0\otimes_\F\overline{\F}$ or the weight space $(\s
\otimes_\F\overline{\F})_1\cong\s_1\otimes_\F\overline{\F}$ is non-zero
(cf.~\cite[p.~33]{Hu}). Let $s$ be a non-zero weight vector in $\s$ of weight 
$0$ or $1$ and define $r:=h\wedge s$. Then by virtue of \cite[Theorem 3.2]{Mi2}, 
$\cyb(r)=0$ and as above it is enough to show that $\overline{\delta_r}\ne 0$. 
Choose $E\in\overline{\lev}$ such that $[\overline{h},E]=2E$. (Note that this 
means in particular that $\overline{h}$ and $E$ are {\it linearly independent\/} 
over $\overline{\F}$.) Then $$\overline{\delta_r}(E)=-2E\wedge\overline{s}
+\overline{h}\wedge [E,\overline{s}]\ne 0,$$ where $\overline{s}:=s\otimes 
1_{\overline{\F}}$. Hence $\sol(\ag)=0$ and thus $\ag$ is three-dimensional 
simple.\quad $\Box$
\bigskip


\section{Quaternionic Lie Bialgebras}


In this section let $\F$ be an arbitrary field of characteristic $\ne 2$. 
In order to characterize those finite-dimensional non-solvable Lie algebras 
which admit non-trivial triangular Lie bialgebra structures, we need to 
consider a certain class of four-dimensional unital associative algebras.

Let $\alpha$, $\beta$ be non-zero elements of $\F$. Then the four-dimensional 
vector space $$(\alpha,\beta)_\F:=\F 1\oplus\F i\oplus\F j\oplus\F k$$ is an 
associative $\F$-algebra with unity element $1$ and the defining relations 
$$i^2=-\alpha\cdot 1,\quad j^2=-\beta\cdot 1,\quad ij=k=-ji.$$ One important 
example are {\it Hamilton's quaternions} which arise as $(1,1)_\R$. Therefore 
any algebra $(\alpha,\beta)_\F$ with $0\ne\alpha,\beta\in\F$ is called a 
{\it quaternion algebra} over $\F$. Note that the assumption $\alpha\ne 0\ne\beta$ 
assures that $i$ and $j$ (and thus also $k$) are not nilpotent. In fact, 
$\alpha\ne 0\ne\beta$ implies that $(\alpha,\beta)_\F$ is a {\it central simple\/}
algebra (cf.~\cite[Lemma 1.6]{P} or \cite[Lemma 11.15 in Chapter 2]{S}).

If $A$ is an associative algebra, then $A$ is also a Lie algebra via the Lie bracket
defined by $[x,y]:=xy-yx$ for every $x,y\in A$ which will be denoted by $\lie(A)$.  
For us the following well-known result will be useful (see \cite[Corollary 1.6.2]{SF}).
\bigskip

\noindent {\bf Lemma 4.} {\it If $\ag$ is a three-dimensional simple Lie algebra 
over an arbitrary field $\F$ of characteristic $\ne 2$, then there exists a 
quaternion algebra $Q$ such that $\ag$ is isomorphic to $\lie(Q)/\F 1_Q$.}
\quad $\Box$
\bigskip

Set $[\alpha,\beta]_\F:=\lie((\alpha,\beta)_\F)/\F 1$ and let $e_1$ denote the 
residue class of $\frac{1}{2}i$, let $e_2$ denote the residue class of $\frac{1}{2}j$, 
and let $e_3$ denote the residue class of $\frac{1}{2}k$ in $[\alpha,\beta]_\F$. The 
Lie algebra $[\alpha,\beta]_\F$ is called a {\it quaternionic Lie algebra\/} over $\F$ 
and we have $$[\alpha,\beta]_\F=\F e_1\oplus\F e_2\oplus\F e_3$$ with the following Lie 
brackets 
\begin{equation}
[e_1,e_2]=e_3,\quad [e_2,e_3]=\beta e_1,\quad [e_3,e_1]=\alpha e_2.
\end{equation}

\noindent {\it Remark.\/} According to $(3)$, every quaternionic Lie algebra is 
perfect which in turn implies that every quaternionic Lie algebra over a field 
$\F$ of characteristic $\ne 2$ is simple (cf.~\cite[3(d), p.~34]{SF}).
\bigskip

A three-dimensional simple Lie algebra $\g$ over $\F$ is called {\it split\/} if 
$\g\cong\slg_2(\F)$ and {\it non-split\/} otherwise. Similarly, a central simple 
algebra over $\F$ is called {\it split\/} if it is isomorphic to $\mat_n(\F)$ for
some positive integer $n$ and {\it non-split\/} otherwise. 
\bigskip

\noindent {\bf Example 1.\/} Note that $(-1,-1)_\F$ is isomorphic to $\mat_2(\F)$
(cf.~the proof of \cite[Corollary 11.14 in Chapter 2]{S}). Let $E_{ij}$ denote the 
$2\times 2$ matrix having a $1$ in the $ij$-th entry and $0$'s otherwise. Moreover,
set $1:=E_{11}+E_{22}$. Then the residue classes $H$, $E$, and $F$ of $E_{11}-E_{22}$, 
$E_{12}$, and $E_{21}$, respectively, in $[-1,-1]_\F=\gl_2(\F)/\F 1$ are linearly 
independent over $\F$ and satisfy the relations $[H,E]=2E$, $[H,F]=-2F$, as well as 
$[E,F]=H$. Consequently, $[-1,-1]_\F\cong\slg_2(\F)$.
\bigskip

\noindent {\bf Example 2.\/} Consider the {\it non-split\/} three-dimensional simple 
real Lie algebra $\su(2)$ (which can be realized as the cross product on three-dimensional 
euclidean space). Then $\su(2)\cong[1,1]_\F$ and it is well-known that $\su(2)$ is 
the only {\it non-split\/} three-dimensional simple real Lie algebra (up to isomorphism). 
Note that this corresponds to Frobenius' classical result that Hamilton's quaternion 
algebra $(1,1)_\R$ is the only non-split central simple $\R$-algebra (up to isomorphism)
(cf.~e.g.~the argument after \cite[Corollary 1.7]{P}). 
\bigskip

The following result determines exactly which quaternionic Lie algebras admit  
non-trivial (quasi-) triangular Lie bialgebra structures.
\bigskip

\noindent {\bf Proposition 1.} {\it Let $[\alpha,\beta]_\F$ be a quaternionic Lie 
algebra over a field $\F$ of characteristic $\ne 2$. Then the following statements 
are equivalent:}
\begin{enumerate}
\item[{\rm(a)}] {\it $[\alpha,\beta]_\F$ admits a non-trivial triangular Lie bialgebra
                structure.}
\item[{\rm(b)}] {\it $[\alpha,\beta]_\F$ admits a non-trivial quasi-triangular Lie
                bialgebra structure.}
\item[{\rm(c)}] {\it $[\alpha,\beta]_\F$ is isomorphic to the split three-dimensional 
                simple Lie algebra $\slg_2(\F)$.}
\end{enumerate}
\smallskip

\noindent {\it Proof\/.} Since the implication (a)$\Longrightarrow$(b) is trivial and 
the implication (c)$\Longrightarrow$(a) is an immediate consequence of \cite[Theorem 
3.2]{Mi2}, it is enough to show the implication (b)$\Longrightarrow$(c).

In order to prove the remaining implication, let us consider the quaternionic Lie algebra
$\g:=[\alpha,\beta]_\F$ with $0\ne\alpha,\beta\in\F$. Then it follows from a straightforward 
computation that $$(\g\otimes\g)^\g=\F(\beta e_1\otimes e_1+\alpha e_2\otimes e_2+e_3
\otimes e_3).$$ Hence, $r\in\g\otimes\g$ satisfies $r+\tau(r)\in(\g\otimes\g)^\g$ if and 
only if $$r=\eta(\beta e_1\otimes e_1+\alpha e_2\otimes e_2+e_3\otimes e_3)+\eta_{12}e_1
\wedge e_2+\eta_{23}e_2\wedge e_3+\eta_{31}e_3\wedge e_1,$$ for some $\eta,\eta_{12},
\eta_{23},\eta_{31}\in\F$. By virtue of \cite[Remark 2 after the proof of Lemma 2.1.3]{CP}, 
we have $$\cyb(r)=\eta^2\cyb(\beta e_1\otimes e_1+\alpha e_2\otimes e_2+e_3\otimes e_3)+
\cyb(\eta_{12}e_1\wedge e_2+\eta_{23}e_2\wedge e_3+\eta_{31}e_3\wedge e_1).$$ A straightforward 
computation yields $$\cyb(\beta e_1\otimes e_1+\alpha e_2\otimes e_2+e_3\otimes e_3)
=(\alpha\beta)e_1\wedge e_2\wedge e_3.$$ Another straightforward computation shows that 
$\{e_1\wedge e_2,e_2\wedge e_3,e_3\wedge e_1\}$ is an {\it orthogonal\/} basis of $\g\wedge\g$ 
with respect to Drinfel'd's Poisson superbracket $\{\cdot,\cdot\}$ (cf.~\cite[Proposition 2]{Fe1}) 
if we identify $\g\wedge\g\wedge\g$ with $\F$ via $e_1\wedge e_2\wedge e_3\mapsto 1_\F$. 
According to \cite[Proposition 2 and (3)]{Fe1}, we obtain 
\begin{equation}
\cyb(\eta_{12}e_1\wedge e_2+\eta_{23}e_2\wedge e_3+\eta_{31}e_3\wedge e_1)=(\eta_{12}^2
+\beta\eta_{23}^2+\alpha\eta_{31}^2)e_1\wedge e_2\wedge e_3.
\end{equation}
Combining the last two results, we conclude that $$\cyb(r)=(\alpha\beta\eta^2+\eta_{12}^2+
\beta\eta_{23}^2+\alpha\eta_{31}^2)e_1\wedge e_2\wedge e_3.$$ Hence $\cyb(r)=0$ if and only 
if $(\eta_{12},\eta_{31},\eta_{23},\eta)\in\F^4$ is an isotropic vector of the quadratic form 
$W^2+\alpha X^2+\beta Y^2+\alpha\beta Z^2$. But the latter is just the norm form of the 
quaternion algebra $(\alpha,\beta)_\F$ and thus \cite[Corollary 11.10 in Chapter 2]{S} 
implies that the CYBE for $\g$ has a non-zero solution with $\g$-invariant symmetric part if 
and only if $(\alpha,\beta)_\F\cong(-1,-1)_\F$. Finally, Example 1 shows that in the latter 
case $[\alpha,\beta]_\F\cong[-1,-1]_\F\cong\slg_2(\F)$.\quad $\Box$
\bigskip

\noindent {\it Remark.\/} It follows from Lemma 4 and Proposition 1 that the classical 
Yang-Baxter equation for a three-dimensional simple Lie algebra $\g$ over a field $\F$ 
of characteristic $\ne 2$ has a non-zero solution with $\g$-invariant symmetric part 
if and only if $\g\cong\slg_2(\F)$. In the latter case, all solutions of the classical 
Yang-Baxter equation with invariant symmetric part are well-known (cf.~\cite{BD} and
\cite[Example 2.1.8]{CP}).
\bigskip

Let us conclude this section by relating the classical Yang-Baxter operator in the 
case of a three-dimensional simple Lie algebra to the determinant. Let $\g$ be a 
three-dimensional simple Lie algebra over a field $\F$ of characteristic $\ne 2$. Then 
the Lie bracket of $\g$ induces a mapping $\gamma$ from $\g\wedge\g$ into $\g$ which is 
bijective since $[\g,\g]=\g$ and $\dim_\F\g=3$. Since in this case $\dim_\F\g\wedge\g
\wedge\g=1$, there is also a canonical bijection $\iota$ from $\g\wedge\g\wedge\g$ onto 
$\F$. In particular, $\iota\circ\cyb$ is a {\it quadratic form\/} with associated symmetric 
bilinear form $\iota\circ\{\cdot,\cdot\}$, where $\{\cdot,\cdot\}$ denotes Drinfel'd's 
Poisson superbracket (cf.~\cite[Proposition 2]{Fe1}).

According to Lemma 4, $\g\cong[\alpha,\beta]_\F$ for some $0\ne\alpha,\beta\in\F$.
Let $\sqrt{\alpha}$ and $\sqrt{\beta}$ denote solutions of $X^2=\alpha$ and $X^2=
\beta$, respectively in the algebraic closure $\overline{\F}$ of $\F$. Moreover, let
$\zeta$ denote a solution of $X^2+1=0$ in $\overline{\F}$. Consider the quaternion 
algebra $(\alpha,\beta)_\F$. Then the mapping defined by
\[
1\mapsto\left(\begin{array}{cc} 1 & 0\\0 & 1\end{array}\right),\quad
i\mapsto\left(\begin{array}{cc} \sqrt{\alpha}\zeta & 0\\0 & -\sqrt{\alpha}\zeta\end{array}\right),\quad
j\mapsto\left(\begin{array}{cc} 0 & \sqrt{\beta}\\-\sqrt{\beta} & 0\end{array}\right),\quad
k\mapsto\left(\begin{array}{cc} 0 & \sqrt{\alpha\beta}\zeta\\\sqrt{\alpha\beta}\zeta & 0\end{array}\right)
\]
defines a {\it two-dimensional faithful\/} representation $\rho$ of $(\alpha,\beta)_\F$ 
over $\overline{\F}$. In fact, $\rho$ is an isomorphism $(\alpha,\beta)_\F\cong
\mat_2(\overline{\F})$ of associative $\F$-algebras. Consequently, the mapping defined by
\[
e_1\mapsto\frac{1}{2}\left(\begin{array}{cc} \sqrt{\alpha}\zeta & 0\\0 & -\sqrt{\alpha}\zeta\end{array}\right),\quad
e_2\mapsto\frac{1}{2}\left(\begin{array}{cc} 0 & \sqrt{\beta}\\-\sqrt{\beta} & 0\end{array}\right),\quad
e_3\mapsto\frac{1}{2}\left(\begin{array}{cc} 0 & \sqrt{\alpha\beta}\zeta\\\sqrt{\alpha\beta}\zeta & 0\end{array}\right)
\]
defines a {\it two-dimensional faithful\/} representation $\rho$ of $\g\cong[\alpha,\beta]_\F$ 
over $\overline{\F}$. An easy calculation shows that $$(\det\circ\rho\circ\gamma)(\eta_{12}e_1
\wedge e_2+\eta_{23}e_2\wedge e_3+\eta_{31}e_3\wedge e_1)=\frac{1}{4}\alpha\beta(\eta_{12}^2+
\beta\eta_{23}^2+\alpha\eta_{31}^2).$$ Comparing this with $(4)$ yields the following result
which was observed for $\slg_2$ in \cite[Example 2.1.8]{CP} and for $\su(2)$ in \cite[Remark 
after Example 1]{Fe1}. 
\bigskip

\noindent {\bf Proposition 2.} {\it Let $\g\cong[\alpha,\beta]_\F$ be a three-dimensional 
simple Lie algebra over an arbitrary field $\F$ of characteristic $\ne 2$ with $0\ne\alpha,
\beta\in\F$. Then the diagram 
\[
\begin{array}{rcccccccl}
\g & \wedge & \g & \stackrel{\cyb}\longrightarrow & \g & \wedge & \g & \wedge & \g\\
\gamma & \downarrow & & & & & \downarrow & \iota &\\
& \g & & \stackrel{\frac{4}{\alpha\beta}\det\circ\rho}\longrightarrow & & & \F & &
\end{array}
\]
is commutative.}\quad $\Box$


\section{Main Results}


Let us consider the three-dimensional {\it solvable\/} Lie algebra 
$$\s_\Lambda(\F)=\F h\oplus\F s_1\oplus\F s_2;$$ $$[h,s_1]=
\lambda_{11}s_1+\lambda_{12}s_2,\quad [h,s_2]=\lambda_{21}s_1+
\lambda_{22}s_2,\quad [s_1,s_2]=0,$$ where $\Lambda:=
(\lambda_{ij})_{1\le i,j\le 2}$ is an element of $\mat_2(\F)$.

If $\Lambda=0$, then $\s_\Lambda(\F)$ is abelian. If $\Lambda\ne 0$ 
is {\it singular\/}, then $\s_\Lambda(\F)$ is either isomorphic to 
the three-dimensional Heisenberg algebra or isomorphic to the (trivial) 
one-dimensional central extension of the two-dimensional non-abelian 
Lie algebra. Finally, if $\Lambda\in\GL_2(\F)$, then $C(\ag)=0$ and 
$\dim_\F[\ag,\ag]=2$. (In fact, $\Lambda\in\GL_2(\F)$ if and only if 
$C(\ag)=0$ if and only if $\dim_\F[\ag,\ag]=2$.) 
\bigskip

\noindent {\it Remark.\/} It is elementary to show that every {\it 
non-simple\/} three-dimensional Lie algebra is isomorphic to 
$\s_\Lambda(\F)$ for a suitable choice of $\Lambda\in\mat_2(\F)$ 
(cf.~e.g.~\cite[Section I.4]{Jac} or \cite[Section 1.6]{SF}).
\bigskip

Let $\F^2:=\{\xi^2\mid\xi\in\F\}$ denote the squares in $\F$. We can 
now state the main result of this paper.
\bigskip

\noindent {\bf Theorem.} {\it Let $\ag$ be a finite-dimensional Lie 
algebra over a field $\F$ of characteristic zero. Then the following 
statements are equivalent:}
\begin{enumerate}
\item[{\rm(a)}] {\it $\ag$ admits a non-trivial triangular Lie bialgebra
                structure.}
\item[{\rm(b)}] {\it $\ag$ admits a non-trivial quasi-triangular Lie
                bialgebra structure.}
\item[{\rm(c)}] {\it $\ag$ is non-abelian and neither a non-split 
                three-dimensional simple Lie algebra over $\F$ nor isomorphic 
                to the three-dimensional Heisenberg algebra $\h_1(\F)$ or  
                $\s_\Lambda(\F)$ with $\tr(\Lambda)=0$ and $-\det(\Lambda)
                \notin\F^2$.}
\end{enumerate}
\bigskip

\noindent {\it Proof\/.} Since the implication (a)$\Longrightarrow$(b) 
is trivial, it is enough to show the implications (b)$\Longrightarrow$(c) 
and (c)$\Longrightarrow$(a).

(b)$\Longrightarrow$(c): According to Lemma 4 and Proposition 1, a non-split 
three-dimensional simple Lie algebra does not admit any non-trivial quasi-triangular 
Lie bialgebra structure. Since it is clear that every coboundary Lie bialgebra 
structure on an abelian Lie algebra is trivial, it suffices to prove that 
$\h_1(\F)$ as well as $\s_\Lambda(\F)$ with $\tr(\Lambda)=0$ and $-\det(\Lambda)
\notin\F^2$ do not admit any non-trivial quasi-triangular Lie bialgebra 
structure. For $\h_1(\F)$ this was already done in \cite[Example 2]{Fe1}. Let 
us now consider $\s:=\s_\Lambda(\F)$ where $\det(\Lambda)\ne 0$. Then a 
straightforward computation yields
\begin{eqnarray*}
(\s\otimes\s)^\s=
\left\{
\begin{array}{cl}
\F(s_1\wedge s_2)\oplus\F[\lambda_{21}(s_1\otimes s_1)-\lambda_{11}(s_1\otimes s_2
+s_2\otimes s_1)-\lambda_{12}(s_2\otimes s_2)] & \mbox{if }\tr(\Lambda)=0\\ 
0 & \mbox{if }\tr(\Lambda)\ne 0
\end{array}
\right.
\end{eqnarray*}
Consider now a $2$-tensor $r=r_0+r_*$ with $\s$-invariant symmetric part
$r_0$ and skew-symmetric part $r_*$. Because of $[s_1,s_2]=0$, we conclude
from \cite[Remark 2 after the proof of Lemma 2.1.3]{CP} that $$\cyb(r)=
\cyb(r_0)+\cyb(r_*)=\cyb(r_*)$$ and $$\delta_r(x)=x\cdot r=x\cdot r_0+
x\cdot r_*=x\cdot r_*=\delta_{r_*}(x)$$ for every $x\in\s$. Consequently,
$\s$ admits a non-trivial quasi-triangular Lie bialgebra structure if and 
only if it admits a non-trivial triangular Lie bialgebra structure. Hence 
it will follow directly from the argument below that $\s$ does {\it not\/} 
admit {\it any\/} non-trivial quasi-triangular Lie bialgebra structure 
unless $\tr(\Lambda)\ne 0$ or $\tr(\Lambda)=0$ and $-\det(\Lambda)\in\F^2$.

(c)$\Longrightarrow$(a): Suppose that $\ag$ does not admit any
non-trivial triangular Lie bialgebra structure. If $\ag$ is not
solvable, then it follows from Lemma 3, Lemma 4, and Proposition 1 
that $\ag$ is isomorphic to a non-split three-dimensional simple 
Lie algebra over $\F$.

If $\ag$ is solvable, then by virtue of Lemmas 1 and 2, we can 
assume for the rest of the proof that the center $C(\ag)$ of $\ag$ 
is zero and $[\ag,\ag]$ is abelian of dimension at most $2$.

Suppose that $\dim_\F[\ag,\ag]=1$. Since $C(\ag)=0$, for any
non-zero element $e\in[\ag,\ag]$ there exists an element $a\in\ag$
such that $[a,e]\ne 0$. (Note that this means in particular that
$a$ and $e$ are {\it linearly independent\/} over $\F$.) But
because of $\dim_\F[\ag,\ag]=1$, we have $[a,e]=\lambda e$ for 
some $0\ne\lambda\in\F$. Set $r:=a\wedge e\in\im(\id_\ag-\tau)$.
Then it follows from \cite[Theorem 3.2]{Mi2} that $r$ is a
solution of the CYBE and $$\delta_r(a)=[a,a]\wedge e+a\wedge [a,e]
=\lambda\cdot(a\wedge e)=\lambda\cdot r\ne 0$$ implies that $\delta_r$ 
defines a non-trivial triangular Lie bialgebra structure on $\ag$.

Hence we can assume from now on that $C(\ag)=0$ and that
$[\ag,\ag]$ is two-dimensional abelian. Since $\dim_\F
[\ag,\ag]=2$, there exist $s_1,s_2\in\ag$ such that $$[\ag,\ag]
=\F s_1\oplus\F s_2.$$

Next, we show that $\dim_\F\ag/[\ag,\ag]=1$. Suppose to the
contrary that $\dim_\F\ag/[\ag,\ag]\ge 2$. Because of $C(\ag)=0$,
there is an element $a\in\ag$ such that $[a,s_1]\ne 0$. In
particular, $a\not\in\F s_1\oplus\F s_2$, i.e., $a$, $s_1$, and
$s_2$ are linearly independent over $\F$. It follows from
$\dim_\F\ag/[\ag,\ag]\ge 2$ that there also is an element
$a^\prime\in\ag$ such that $a$, $a^\prime$, $s_1$, and $s_2$ are
linearly independent over $\F$. Moreover, for every $1\le i,j\le
2$, there exist elements $\alpha_{ij},\alpha_{ij}^\prime\in\F$
such that
\begin{eqnarray*}
&& [a,s_1] \; = \alpha_{11}s_1+\alpha_{12}s_2,\quad [a,s_2] \; =
\alpha_{21} s_1+\alpha_{22}s_2,\\ && [a^\prime,s_1] =
\alpha_{11}^\prime s_1+ \alpha_{12}^\prime s_2,\quad
[a^\prime,s_2] = \alpha_{21}^\prime s_1 +\alpha_{22}^\prime s_2.
\end{eqnarray*}
If $\alpha_{12}=0$, then $[a,s_1]=\alpha_{11}s_1\ne 0$, and one
can argue as above (for the case $\dim_\F[\ag,\ag]=1$) that $\ag$
admits a non-trivial triangular Lie bialgebra structure. On the
other hand, if $\alpha_{12}\ne 0$, let us set $h:=\alpha_{12}^\prime 
a-\alpha_{12}a^\prime$ and $\lambda:=\alpha_{12}^\prime\alpha_{11}-
\alpha_{12}\alpha_{11}^\prime$. Then $h\not\in[\ag,\ag]$ and $[h,s_1]
=\lambda s_1$. If we now put $r:=h\wedge s_1\in\im(\id_\ag-\tau)$, we
see as before that $r$ is a solution of the CYBE. Since $h\not\in 
[\ag,\ag]$ and $[a,s_1]\ne 0$, we conclude that $$\delta_r(a)=[a,h]
\wedge s_1+h\wedge[a,s_1]\ne 0.$$ Hence $\delta_r$ defines a non-trivial
triangular Lie bialgebra structure on $\ag$.

Finally, we can assume that $\ag$ is three-dimensional and
$[\ag,\ag]$ is two-dimensional abelian. It follows that $\ag
\cong\s_\Lambda(\F)$ with $\det(\Lambda)\ne 0$. Then we obtain 
for an arbitrary skew-symmetric $2$-tensor $$r=\omega s_1\wedge 
s_2+\xi_1 h\wedge s_1+\xi_2 h\wedge s_2$$ with $\omega,\xi_1,
\xi_2\in\R$ that $$\cyb(r)=[\lambda_{12}\xi_1^2-(\lambda_{11}-
\lambda_{22})\xi_1\xi_2-\lambda_{21}\xi_2^2]\cdot h\wedge s_1
\wedge s_2.$$ If $\tr(\Lambda)\ne 0$, then -- as already 
established in the proof of the implication (b)$\Longrightarrow$(c) 
-- there is {\it no\/} non-zero $\s_\Lambda(\F)$-invariant 
$2$-tensor. Consequently, $r:=s_1\wedge s_2$ defines a non-trivial 
triangular Lie bialgebra structure on $\s_\Lambda(\F)$.

On the other hand, if $\tr(\Lambda)=0$, then the discriminant of
the relevant homogeneous quadratic equation $$\lambda_{12}X_1^2
-(\lambda_{11}-\lambda_{22})X_1X_2-\lambda_{21}X_2^2=0$$ is the 
{\it negative\/} of $\det(\Lambda)$. Hence $\s_\Lambda(\F)$ admits 
a non-trivial triangular Lie bialgebra structure if and only if 
$-\det(\Lambda)\in\F^2$.\quad $\Box$
\bigskip

\noindent As an immediate consequence of the theorem we obtain the
following existence result:
\bigskip 

\noindent {\bf Corollary 1.} {\it If $\ag$ is a finite-dimensional 
non-abelian Lie algebra over a field $\F$ of characteristic zero 
with $\dim_\F\ag\ne 3$, then $\ag$ admits a non-trivial triangular 
Lie bialgebra structure.}\quad $\Box$
\bigskip

We conclude with the following generalization of the main result
of \cite{dSm} from $\R$ and $\C$ to arbitrary ground fields of
characteristic zero (for another generalization see also the 
remark after \cite[Theorem 4]{Fe1}).
\bigskip

\noindent {\bf Corollary 2.} {\it Every finite-dimensional non-abelian 
Lie algebra over an arbitrary field $\F$ of characteristic zero admits 
a non-trivial coboundary Lie bialgebra structure.}
\bigskip

\noindent {\it Proof\/.} According to the theorem, we only have to prove 
the existence of a non-trivial coboundary Lie bialgebra structure for a 
non-split three-dimensional simple Lie algebra $\g$ over $\F$, the 
three-dimensional Heisenberg algebra $\h_1(\F)$, and the three-dimensional 
solvable Lie algebra $\s_\Lambda(\F)$ with $\tr(\Lambda)=0$ and 
$-\det(\Lambda)\notin\F^2$.

First, let us consider a non-split three-dimensional simple Lie algebra $\g$ 
over $\F$. By virtue of Lemma 4, $\g$ is a quaternionic Lie algebra $[\alpha,
\beta]_\F$ for some $0\ne\alpha,\beta\in\F$. Set $r:=e_1\wedge e_2$. Then it 
follows from \cite[Proposition 2]{Fe1} that $\cyb(r)=2e_1\wedge e_2\wedge e_3$ 
and thus a straightforward computation shows that $\cyb(r)$ is $\g$-invariant. 
Because of $\delta_r(e_1)=2e_1\wedge e_3\ne 0$, the skew-symmetric $2$-tensor 
$r$ defines a non-trivial coboundary Lie bialgebra structure on $\g$.

In the case of the three-dimensional Heisenberg algebra $\h:=\h_1(\F)$ set $r:=
p\wedge q$. It was shown in the proof of \cite[Theorem 3]{Fe1} that $\cyb(r)=p
\wedge q\wedge\hbar$ which clearly is $\h$-invariant. Since $$\delta_r(p)=p
\wedge\hbar\ne 0\ne q\wedge\hbar=\delta_r(q)\,,$$ the skew-symmetric $2$-tensor 
$r$ defines a non-trivial coboundary Lie bialgebra structure on $\h$.

Finally, consider $\s:=\s_\Lambda(\F)$ with $\tr(\Lambda)=0$ and $-\det(\Lambda)
\notin\F^2$. Set $r_1:=h\wedge s_1$ and $r_2:=h\wedge s_2$. Then it follows from 
\cite[Proposition 2]{Fe1} that $\cyb(r_1)=\lambda_{12}h\wedge s_1\wedge s_2$ and
$\cyb(r_2)=-\lambda_{21}h\wedge s_1\wedge s_2$. But obviously, $s_1\cdot(h\wedge
s_1\wedge s_2)=0$, $s_2\cdot(h\wedge s_1\wedge s_2)=0$, and $h\cdot(h\wedge s_1
\wedge s_2)=\tr(\Lambda)h\wedge s_1\wedge s_2$. Hence $\tr(\Lambda)=0$ implies 
that $\cyb(r_1)$ and $\cyb(r_2)$ are both $\s$-invariant. On the other hand,
$$\delta_{r_1}(h)=\lambda_{11}h\wedge s_1+\lambda_{12}h\wedge s_2$$ and $$
\delta_{r_2}(h)=\lambda_{21}h\wedge s_1+\lambda_{22}h\wedge s_2$$ show that at
least one of the skew-symmetric $2$-tensors $r_1$ and $r_2$ defines a non-trivial 
coboundary Lie bialgebra structure on $\s$ if $\tr(\Lambda)=0$ and $\Lambda\ne 0$.
\quad $\Box$
\bigskip






\end{document}